\documentclass[twocolumn,showpacs,aps,prx]{revtex4-2}
\usepackage{chngcntr}

\usepackage{amssymb}
\usepackage{amsmath}
\usepackage{graphicx}
\usepackage{bm}
\usepackage{epstopdf}
\usepackage{hyperref}
\usepackage[all]{hypcap}
\hypersetup{
    colorlinks=true,
    linkcolor=blue,
    filecolor=magenta,
    urlcolor=cyan,
    citecolor=red
     }

\begin{document}

\title{Acceleration and focusing of multispecies ion beam using a converging laser-driven shock}

\author{Jihoon Kim$^1$, Roopendra Rajawat $^1$, Tianhong Wang$^1$, Gennady Shvets$^1$}
\affiliation{$^1$School of Applied and Engineering Physics, Cornell University, Ithaca, NY 14850, USA.
}

\begin{abstract}
 We demonstrate an ion acceleration scheme capable of simultaneously focusing and accelerating a multispecies ion beam with monoenergetic spectra to a few micron radius. The focal length and ion mean energy can be independently controlled: the former by using a different front-surface shape and the latter by tuning the laser-plasma parameters. We interpret the results using simple models and validate the results using first-principles simulations. The scheme is applicable to different laser transverse profiles and multi-ion species target, and limiting factors for the ion focusing are delineated. The generated ion beam exhibits high charge, low emittance, and high energy flux and is of interest to various applications including Inertial Confinement Fusion (ICF), high flux neutron generation, and biomedical applications.

\end{abstract}

\maketitle


\section{Introduction}
Chirped Pulse Amplification (CPA) \cite{CPA} has made investigation of ultra-intense light-matter interaction possible, giving rise to multi-petawatt laser systems \cite{PW_laser}. One promising application for these high-power laser systems is ion acceleration \cite{Macchi, Zhang_review, Eskirepov_2004,Naumova_prl,Naumova_POP,Robinson_PPCF}. The laser can impart energy to the over-dense electrons, separating them from the bulk ions at the front   \cite{Naumova_POP,Naumova_prl,Robinson_PPCF,Eskirepov_2004,Macchi_HB} or the rear surface \cite{Wilks_TNSA,Patel,Pukhov_PRX,Hegelich_BOA}. The resulting charge-separation field can then accelerate ions to tens of MeV within in a few microns, giving name to the “table-top ion accelerator”. The laser pulses accelerating these ion beams are ultrashort (femto- to picosecond) and often tightly focused (tens of microns), which in turn translates to ion beams with comparable dimensions.

Some of the most promising applications for ion beams require energy flux higher than those available by current technology. In particular, Fast Ignition (FI) \cite{Tabak,Fernandez,Honrubia_carbon,Honrubia_carbon2} of Inertial Confinement Fusion (ICF) \cite{NIF_Nature, NIF_Nature_physics,NIF_PRL} targets require delivery of energy flux in the order of $\mathrm{GJ/cm^2}$ to heat the compressed fusion fuel and reach net energy gain \cite{Naumova_prl, Atzeni_1999}. 
 While there have been proposals to generate such ion beams via extremely energetic and powerful (45kJ, 60PW) lasers \cite{Naumova_prl}, such lasers are yet unavailable \cite{PW_laser}. 
 
One way to attain the high energy flux without increasing laser energy and power is by focusing the ions. Several methods have been proposed: combination of hemispherical target rear shape and Target Normal Sheath Acceleration (TNSA) with or without a guiding cone \cite{Patel, Beg,Bhutwala}, post-acceleration focusing of ion beams using a cylindrical expanding sheath \cite{Toncian}, and using a parabolic density target to fold an ultra-thin target to a singular focal spot \cite{LILA,Roopendra}. 

These schemes have their respective advantages and disadvantages, often requiring additional manipulation to generate beams that meet the application's requirement. In particular, TNSA relies on generation of hot electrons which in turn leads to ions with exponential energy spectra \cite{Wilks_TNSA}. Furthermore, it primarily accelerates lighter ions such as protons, and often have quite low energy conversion from laser to ions. The post-acceleration focusing of ion beams using a cylindrical expanding sheath requires a separate laser and enough space to implement the focusing mechanism.  While the self-folding target yields monoenergetic well-focused ions, this method generates highly relativistic ions, thus limiting its applicability.


 \begin{figure}[t]
    \includegraphics[width=0.5 \textwidth]{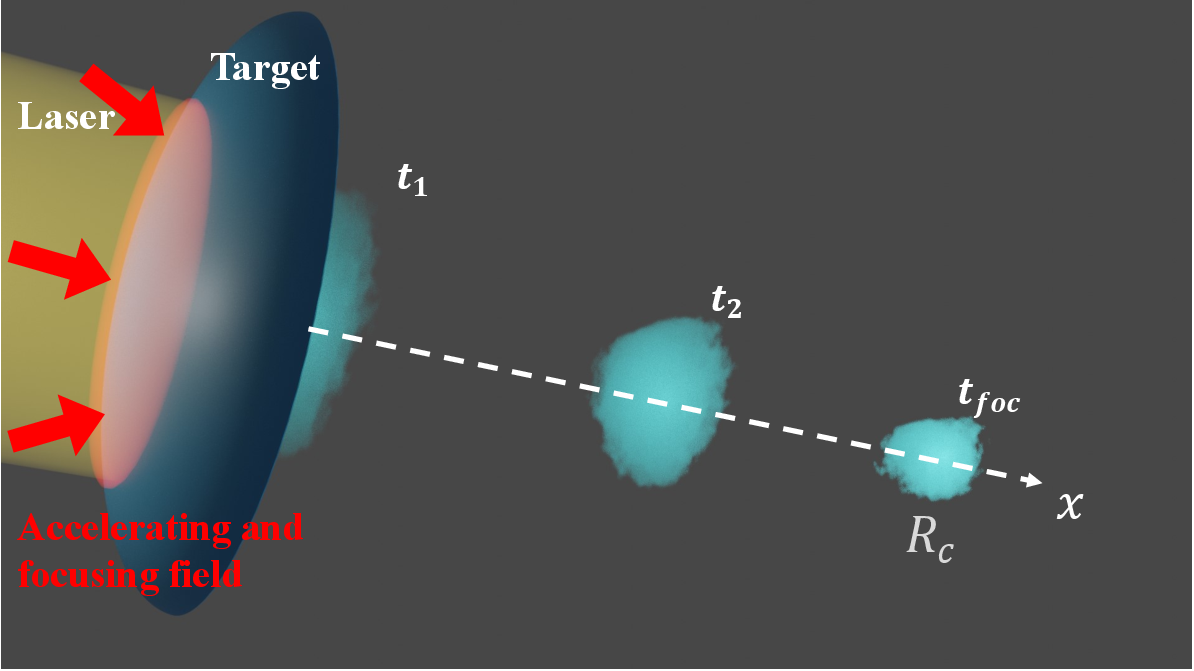}
\caption{CLIA Schematics. A Circularly polarized laser (yellow) is incident on an overdense target with front side shaped like a paraboloid (dark blue), generating an electric field (red), accelerating and focusing (red thick arrows) a quasi-neutral plasma (light blue). After the initial acceleration, the beam propagates almost ballistically, forming a high energy flux ion beam at focal length $R_c$ at time $t_{\rm foc}$ }\label{fig:schematic}
\end{figure}

In this paper, we propose a mechanism to accelerate ions with monoenergetic spectra with arbitrary mean energy to a predefined focal length. We combine the Hole Boring-Radiation Pressure Acceleration (HB-RPA) mechanism \cite{Robinson_PPCF,Naumova_POP,Macchi_HB} and front target shaping to simultaneously focus and accelerate the ion beams into a small spot at the focal length, forming a Converging-front Laser Ion Accelerator (CLIA) [Fig \ref{fig:schematic}]. Two key mechanisms make CLIA possible. First, a planar Circularly Polarized (CP) laser interacting with a thick over-dense plasma target can maintain the Plasma-Laser Interface (PLI) shape   throughout the HB-RPA process. Second, the electric field from the parabolic front surface of such PLI impart ions with transverse velocity proportional to the distance away from the axis, focusing them simultaneously to a predefined distance $R_c$. These two effects generate monoenergetic beam whose energy flux can be enhanced by an order of magnitude in comparison to that obtained from a flat target. Furthermore, the peak beam energy and focal length can be independently controlled: the former by tuning the target and laser parameters and the latter by shaping target front surface. 

This paper is organized as follows. In section \ref{section:3D_PIC},  we present a 3D Particle-In-Cell (PIC) simulation demonstrating CLIA. In section \ref{sec:theory}, we interpret the result using models that explain the laser-plasma interface evolution, the ion acceleration process, and the subsequent ion propagation in vacuum.  We compare the theoretical descriptions with field structure and particle data obtained from the PIC simulation. In section \ref{sec:expt} we use several different laser intensity, focal length, target density, and target geometry to show that CLIA mechanism is applicable to a wide range of experimental parameters, showing the applicability and limits of CLIA. In section \ref{sec:discussion}, we discuss the effect of multispecies target and non-planar laser profile on CLIA scheme. 

\section{Results}
\subsection{3D PIC simulation}\label{section:3D_PIC} 
 \begin{figure}[h]
    \includegraphics[width=0.5 \textwidth]{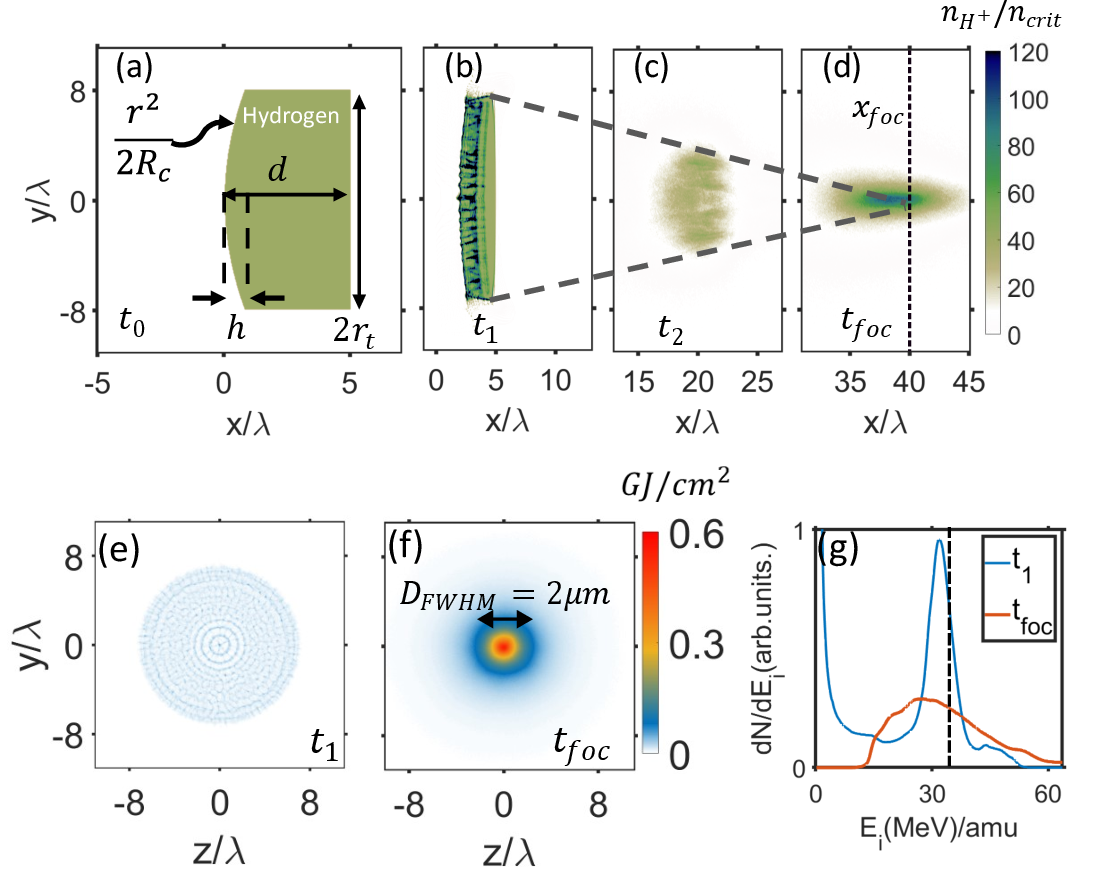}
\caption{3D PIC Simulation results. (a-d): Evolution of proton density in x-y plane at different times. (a) initial proton density at $t_0=0$ with relevant dimensions $h, d, r_t, R_c$ defined. $R_c=40\mathrm{\mu m}$: focal length, $r_t=8\mathrm{\mu m}$ : transverse target radius, $h=r_t^2/(2R_c)\approx 0.9\mathrm{\mu m}$: thickness of the curved part of target,  $d=5\mathrm{\mu m}$: thickness at the center of target. (b-d): proton density immediately after acceleration at $t_1= 50 {\rm fs}$ (b), during focusing at $t_2= {\rm 267}$fs (c), and at focus $t_{\rm{foc}}= {\rm 483}$fs (d). Gray dashed lines in (c-d): guide to the eye showing proton focusing, Black Dashed line in (d): $x_{\rm foc}=R_c$. (e-f): ion energy flux $J=dE_{i}/dA$  at $t_1$ (e) and $t_{\rm foc}$ (f). Black arrow in (f): energy flux FWHM. (g) Energy spectra of all ions right after acceleration at $t_1 \approx 50$fs (blue) and accelerated ions at focus $t_{\rm foc}= 483$fs (red). Black dashed line: $E_i^{\rm th} \approx 35$MeV.  Simulation details: see table \ref{table:1}   }\label{fig:3D PIC}
\end{figure}

We use Smilei \cite{Smilei}, a first-principles 3D PIC code, to self-consistently model the overdense plasma target acceleration and focusing by a Circularly Polarized (CP) laser. A  planar CP laser with wavelength $\lambda_L=1\mathrm{\mu m}$ is incident on a hydrogen target with $n/n_{\rm crit}=40$, or mass density of $0.074\mathrm{g/cm^3}$, close to liquid/ solid hydrogen density. Here, $n_\mathrm{crit}=m_e\omega_L^2/4\pi e^2$, with $m_e$ the electron mass, $\omega_L$ the angular laser frequency, and $e$ the electron charge. The target has a curved front surface defined by $x=r^2/2 R_c$ with $R_c=40\mathrm{\mu m}$, transverse radius $r_t=8\mathrm{\mu m}$, center thickness $d=\mathrm{5\mu m}$, and front bulge with thickness $h=r_t^2/(2 R_c)\approx 0.9\mathrm{\mu m}$ [Fig \ref{fig:3D PIC} (a)].

The laser has intensity $I=5\times 10^{21} \mathrm{W/cm^2}$ ($a_0=43=E_L/E_0$, with $E_L$ the amplitude of transverse laser field and $E_0=m_e c\omega_L/e$, with $c$ the speed of light), and power $P=10\mathrm{PW}$ incident on target. The laser longitudinal profile consists of linear up-downramp of $\tilde{t}_{\rm up}=\tilde{t}_{\rm down}=t_{\rm up}/T_0= t_{\rm down}/T_0=1 (t_{\rm up}=t_{\rm down}=3.3{\rm fs})$ surrounding a $\tilde{t}_{\rm flat}=t_{\rm flat}/T_0=11 (t_{\rm flat}= 36.7{\rm fs})$ flat-top region, with total energy of 400J deposited onto the target, and $T_0=\lambda_L/c$ the single-cycle duration of the laser. We refer to this simulation as Sim I throughout the paper (see table \ref{table:1} for detailed simulation parameters). 

During the initial stage of interaction, the laser pushes on the convex PLI, establishing an interface that accelerates and focuses the ions steadily while moving at a constant velocity $v_b\approx 0.14c$ [Fig \ref{fig:3D PIC} (b)]; the PLI velocity is in agreement with our model (Sec. \ref{sec:theory}). The ions are accelerated to form a monoenergetic spectra with a peak at $E_i\approx35\mathrm{MeV}$, in good agreement with theoretical prediction (Sec. \ref{sec:theory}). The convex shape of the PLI is preserved for almost $2\mathrm{\mu m}$ of target shortly after the interaction is terminated at $t_1=50\mathrm{fs}$ after the entire laser beam duration is utilized for ion acceleration and focusing. The accelerated ions pass through the un-shocked part of the target during this stage, as evidenced by the ion density increase. We note the  transverse striations visible on the PLI [Fig \ref{fig:3D PIC} (b)] arising from the front-surface laser-plasma instability \cite{Yan_RTI_2020,Mori_RTI_2015,Mori_RTI_2018,Macchi_RTI_2015,Eliasson_RTI_2015,Vladimir_RTI_2014}. 

The accelerated ions and the accompanying electron cloud co-propagate as a quasi-neutral plasma after the initial acceleration,  contracting transversely and dilating longitudinally [Fig \ref{fig:3D PIC} (b-d)]. After $t_{\rm foc}= 483$fs, the ion beam focuses at distance $x_{\rm foc}=40\mathrm{\mu m}=R_c$. This agreement between the target curvature $R_c$ and ion focal length $x_{\rm foc}$ can be well explained by theory [Sec. \ref{sec:theory}]. The peak ion density increases to three times the initial ion density at focus despite the beam's longitudinal dilation. We note that the transverse instability visible during the initial stage disappears during the focusing process due to merging of the different filaments. 

The ion beam energy flux increases, reflecting the effect of focusing. At $t_1$, right after the acceleration, the ion beam energy flux shows concentric transverse pattern arising from front surface transverse instability, with the average flux of the ion beam around $0.05\mathrm{GJ/cm^2}$ [Fig \ref{fig:3D PIC} (e)]. The quasineutral beam subsequently propagates and focuses in vacuum, with the transverse striations completely disappearing and the peak energy flux increasing to almost $0.6\mathrm{GJ/cm^2}$ [Fig \ref{fig:3D PIC} (f)], with FWHM of $2\mathrm{\mu m}$. We note that the initially sharp monoenergetic ion peak deteriorates to a wider distribution with the same center at 35MeV due to interaction with the neutralizing hot electron cloud [Fig \ref{fig:3D PIC} (g)].

\subsection{Theoretical Description}\label{sec:theory}

\begin{figure}[ht]
    \includegraphics[width=0.5 \textwidth]{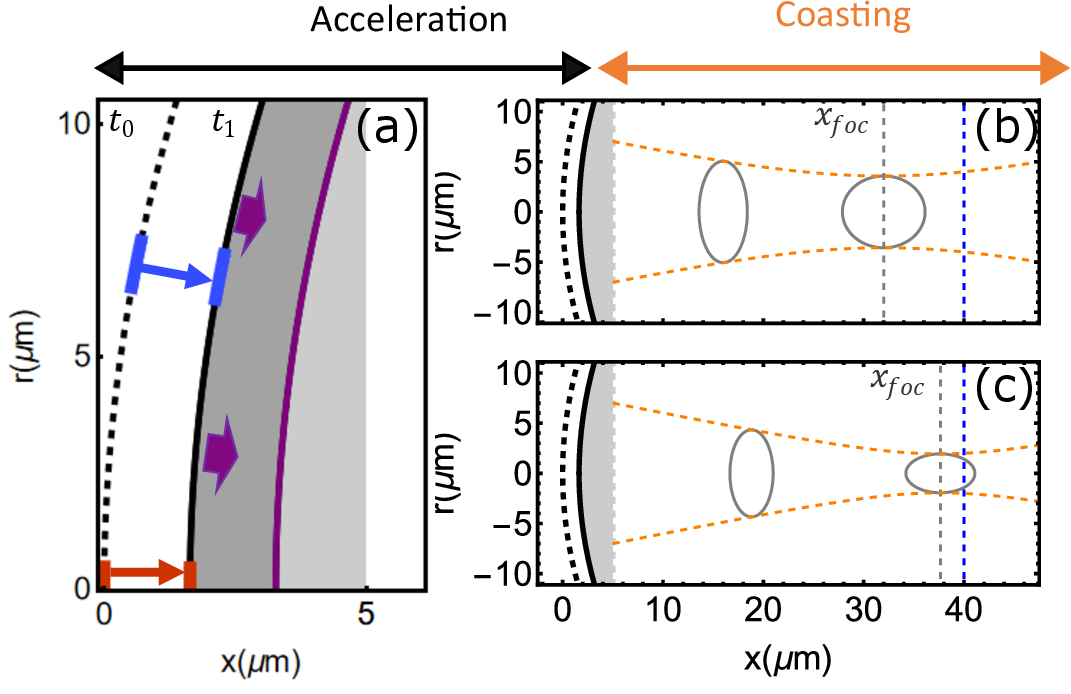}
\caption{Theoretical Description of CLIA. (a) Acceleration stage of CLIA. Blue and Red thick lines and arrows: evolution of different surface elements from $t_0=0$ to $t_1=12\lambda_L/c$ on axis (Red) and off axis (Blue), Black Dashed line (Black solid line): PLI at $t_0$ ($t_1$). Purple arrows: direction of ion acceleration at $t_1$, Gray: ion density, Purple Solid line: position of the leading ions accelerated right after $t_0$ at $t_1$. (b-c) Quasineutral beam propagation in vacuum for two different temperature $T_{e1}=1.72m_ec^2\approx880\mathrm{keV}$(b) and $T_{e2}=T_{e1}/4\approx 220\mathrm{keV}$(c). Gray shaded area, black solid/dashed lines: un-accelerated part of the target as shown in (a) at the end of the acceleration process.  Gray ellipses: quasineutral plasma beam boundary at $t=t_{\rm foc}/2, t_{\rm foc}$. Orange Dashed Lines: outer boundary of quasineutral plasma, Gray dashed line: $x_{\rm foc}=v_i t_{\rm foc}$, Blue Dashed line: $x_{\rm foc}^{\rm bal}=R_c$ for $T_e=0$. Parameters: $R_c=40\mathrm{\mu m}$, $\Theta=0.025$ ($E_i=35 \rm{MeV}$), $T_{e1}\approx880\rm{keV}$(b), $T_{e2}\approx220\rm{keV}$ (c) }\label{fig:theory}
\end{figure}

To interpret CLIA, we adopt simple models based on HB-RPA \cite{Naumova_POP,Naumova_prl,Robinson_PPCF} and quasineutral plasma evolution in vacuum \cite{Bellei,Kovalev,Semenov,Mora,Mora_pop}.  CLIA consists of two distinct stages: Acceleration and Coasting. During the acceleration stage, constant stream of ions are generated at the parabolic front surface with radially varying focusing velocity [Fig \ref{fig:theory} (a)]. Afterwards, the accelerated ion beam propagates through the pristine part of the target into the vacuum until it reaches the focal spot with an accompanying electron cloud; we refer to this as the coasting stage [Fig \ref{fig:theory} (b-c)]. 

We model the acceleration stage by extending the Hole-Boring RPA (HB-RPA) theory. First, we summarize the 1-D HB-RPA mechanism \cite{Naumova_POP,Naumova_prl,Robinson_PPCF}: when an intense Circularly Polarized laser with intensity $I$ is normally incident on a thick overcritical plasma with mass density $\rho_0$, the plasma electrons are pushed forward via the radiation pressure of the laser, leaving the heavier ions behind. Electrostatic shock forms at the front PLI due to charge separation, moving longitudinally with almost constant velocity and accelerating a steady stream of ions.

The acceleration after PLI formation can be considered as a quasi-static process where the outgoing ion momentum flux is balanced by the incoming radiation pressure at the PLI frame. A single parameter $\Theta=I/\rho_0 c^3$, or alternatively,  $\Theta=a_0^2/\left\{(m_i/m_e)(n_e/n_{\rm crit})\right\}$, characterizes the process, with  $m_i$ the  ion mass,  and  $n_e$ the plasma density.  The PLI propagates at velocity $v_{b0}=c\beta_{b0}=c\sqrt{\Theta}/(1+\sqrt{\Theta})$, generating a constant stream of ion beams with velocity $v_{i0}=2v_{b0}/(1+\beta_{b0}^2)$ and energy $E_{i}/m_{i}c^2=2\Theta/(1+2\sqrt{\Theta})$.

Since CLIA involves both transverse and longitudinal motion of ions, we extend the HB-RPA theory to  a cylindricaly symmetric system with two dimensions, $r$ and $x$. When the laser is incident on the surface at an angle $\phi_i$, the radiation pressure on the PLI moving with $\beta_b = v_b/c$ at an angle $\phi_i$ is
$P_{\rm rad}=(2I/c)(\cos\phi_i-\beta_b)^2/(1-\beta_b^2)$\cite{Einstein}. Equating radiation pressure $P_{\rm rad}$ and the outgoing ion momentum, $P_{\rm out}=\gamma_i (m_i n_i v_b )v_i$ with $v_i =2v_b/(1+\beta_b^2)$, the normalized interface velocity $\beta_b$ can be algebraically solved for: 
\begin{equation}
    \beta_b=\beta_{b0}\cos \phi_i \label{eq:HB}.
\end{equation}

A surface element located at $(x,r)$ moves with velocity
\begin{flalign}
    \mathbf{v_b}(r)&=\left(\frac{\partial x}{\partial t}\vec{e_x},\frac{\partial r}{\partial t}\vec{e_r}\right)\label{eq:HB_eqn}\\
    &=v_{b0} \cos \phi_i(r) \left\{ \cos \phi_i(r)\vec{e_x},  \sin \phi_i(r)\vec{e_r}\right\}\nonumber 
\end{flalign} 
and accelerates the ions to speed $\beta_i=v_i/c=2\beta_b/(1+\beta_b^2)$, with $x$ the longitudinal coordinate, $r$ the radial coordinate, and $\vec{e_x}$, $\vec{e_r}$ the longitudinal and radial unit vectors.

We first show that the PLI interacting with a plane CP laser can maintain its shape throughout the interaction. PLI is described by $x(r,t)$, with its evolution written as $dx(r,t)/dt=\partial_r x \, \partial_t r+\partial_t x$. Using Eqn. \ref{eq:HB_eqn}, we find that the evolution of PLI is described by the following equation: 
\begin{equation}
    \frac{dx(r,t)}{dt}=v_{b0}(r,t) \label{eqn:surface_evolution}.
\end{equation}
Even though the PLI undergoes evolution in the $(x,r)$ plane, the net evolution is described by $v_{b0}(r,t)$, which in turn only depends on the instantaneous laser intensity at time $t$ and radial position $r$. We note that this is a general result applicable to  laser with radial spatial dependence interacting with an overdense plasma boundary.

The PLI shape can be described throughout the interaction by Eqn \ref{eqn:surface_evolution} [Fig. \ref{fig:theory} (a), black lines]. In particular, if the laser transverse intensity throughout the entire target is constant (planar wave), the PLI moves at a constant longitudinal velocity $v_{b0}$ throughout the interaction [Fig. \ref{fig:theory} (a)] maintaing the initial shape. The ions are accelerated perpendicular to PLI surface with velocity 

\begin{equation}
    \mathbf{v_i}(r)
    =v_{i}\left\{ \cos \phi_i(r)\vec{e_x}, -\sin \phi_i(r)\vec{e_r}\right\}.\label{eqn:ion_vel} 
\end{equation}
with a transversely focusing velocity component . These ions move ahead of the PLI [Fig \ref{fig:theory}(a), Dark shaded region] with approximately twice the PLI speed \cite{Robinson_PPCF}.

Using the above result, we show that using a parabolic PLI can focus the ions at a well-defined focal length. The incidence angle $\phi_i$ for a parabolic PLI defined by $x=r^2/(2R_c)$ is given by $\phi_i=\arctan{\left(r/R_c\right)}$. For $r\ll R_c$ the small angle approximation, $\phi_i\ll 1$, can be applied, and the ion velocity can be estimated by

\begin{equation}
    \mathbf{v_{i}}(r)\approx \frac{2 v_{b0}}{1+\beta_{b0}^2} \left(\vec{e_x},-\frac{r}{R_c}\vec{e_r}\right).\label{eqn:vion_parax}
\end{equation}

Under this approximation, different PLI elements propagate with a transverse angle given by $\phi_i=v_r/v_x=-r/R_c$. Assuming ballistic ion beam propagation, all the different elements focus on axis at time 
\begin{equation}
    t_{\rm foc}^{\rm bal}=r/{(\mathbf{v_{i}}\cdot \vec{e_r})}\approx \frac{R_c}{2 v_{b0}}(1+\beta_{b0}^2)
\end{equation} and focal length
\begin{equation}
    x_{\rm foc}^{\rm bal}=t_{\rm foc}\times  {(\mathbf{v_i}\cdot \vec{e_x})}=R_c.
\end{equation}, explaining the coincidence of focal length $x_{\rm foc}$ and front surface curvature $R_c$ in section \ref{section:3D_PIC}.
The kinetic ion energy $E_i$ differs negligibly from the 1D theoretical estimate, and the $\Theta$ parameter from the 1-D theory can be used to estimate the ion energy. For example, the monoenergetic ion beam energy is well estimated using $\Theta=0.025$ in $E_i/(m_{i} c^2)=2\Theta/(1+2\sqrt{\Theta}) \simeq35 \rm MeV$ [Fig \ref{fig:3D PIC} (g)].

After the acceleration, the ions are accompanied by a neutralizing electron cloud. When the ions exit the plasma slab and propagate in vacuum, they can be considered as a  cold ion beam accompanied by a hot electron cloud in the coasting stage. Analytic description of such quasineutral plasma evolution in vacuum have been studied by several authors \cite{Kovalev,Semenov,Bellei,Mora,Mora_pop}. In particular, an adiabatically evolving quasineutral electron-ion plasma's characteristic length scales $l_k(t_{a})$  can be described in the nonrelativistic limit \cite{Semenov,Kovalev,Mora_pop}, with $t_{a}$ the time of plasma adiabatic evolution (i.e. after the quasineutral plasma stops interacting with any external electromagnetic forces such as the shock front). We reproduce the relevant equations from \cite{Semenov} below:
\begin{equation}
    l_k^2(t_a)=[l_k(0)+\dot{l}_k(0)t_{a}]^2+c_k^2t_{a}^2.
\end{equation} Here, $c_k^2=(Z m_e V^e_k(0)^2+ m_i V^i_k(0)^2)/(m_i+Zm_e)$  defines the ion acoustic velocity $c_k$ in the $k_\mathrm{th}$ direction, with $l_k$ the characteristic length-scale in $k_{\rm th}$ direction, $m_i$ the ion mass, and $V^{(i,e)}_k(t_a)$ the ion/electron root mean square velocity in the $k_{\rm th}$ direction, and $Z$ the ion charge.

CLIA generates ion beams with a single well defined $\dot{l}_r(0)$ since the contracting velocity is proportional to the distance away from axis. We first note that if the plasma is cold, i.e. $c_k=0$, the ions move ballistically such that the quasineutral plasma beam can contract to almost a singular point at time $t^{\rm bal}_{\rm foc}=l_r(0)/\dot{l}_r(0)$. For finite-temperature plasma, the time $t_{\rm foc}$ to contract to minimum transverse size $l_r^{\rm min}$ are as follows: 

\begin{flalign}
        &l^{\rm min}_r= \frac{c_k l_r(0)}{\sqrt{c_k^2+\dot{l}_r(0)^2}}=c_k  \sqrt{-\frac{l_r(0)}{\dot{l}_r(0)}t_{\rm foc}},
        \\ &t_{\rm foc}=-\frac{l_r(0) \dot{l}_r(0)}{c_k^2+\dot{l}_r^2(0)}.\label{eqn:focus}
\end{flalign}

For visualization, we consider an ellipsoidal plasma longitudinally contracting as it propagates. The plasma moves with longitudinal velocity $v_x(0)/c=0.25$ with $l_x=1\mathrm{\mu m}$, $l_r(0)=8\mathrm{\mu m}$, $\dot{l}_r(0)=v_i(0) l_r/R_c$, $\dot{l}_x(0)=0.05 v_x$, with $R_c=40\mathrm{\mu m}$. The plasma is comprised of cold hydrogen ions and electrons with different temperatures. We have chosen parameters comparable to that from Sim I. Specifically, for the result in Fig \ref{fig:theory} (b), we prescribed $T_{e1}=880\rm keV$ which was extracted from Sim I. The plasma contracts maximally at a shorter focal length with larger transverse radius of $l_r =3.7\mathrm{\mu m}$ [Fig \ref{fig:theory}(b)], forming an elongated ellipse as observed in Sim I [Fig \ref{fig:3D PIC}]. We note if the electron temperature is lower ($T_{e2}=220\rm keV$), the plasma focuses at a longer focal length with smaller radius of $l_r =1.7\mathrm{\mu m}$ [Fig \ref{fig:theory} (c)]; the lower thermal pressure of the electrons can lead to better focusing of the beam.  

\begin{figure}[h]
    \includegraphics[width=0.5 \textwidth]{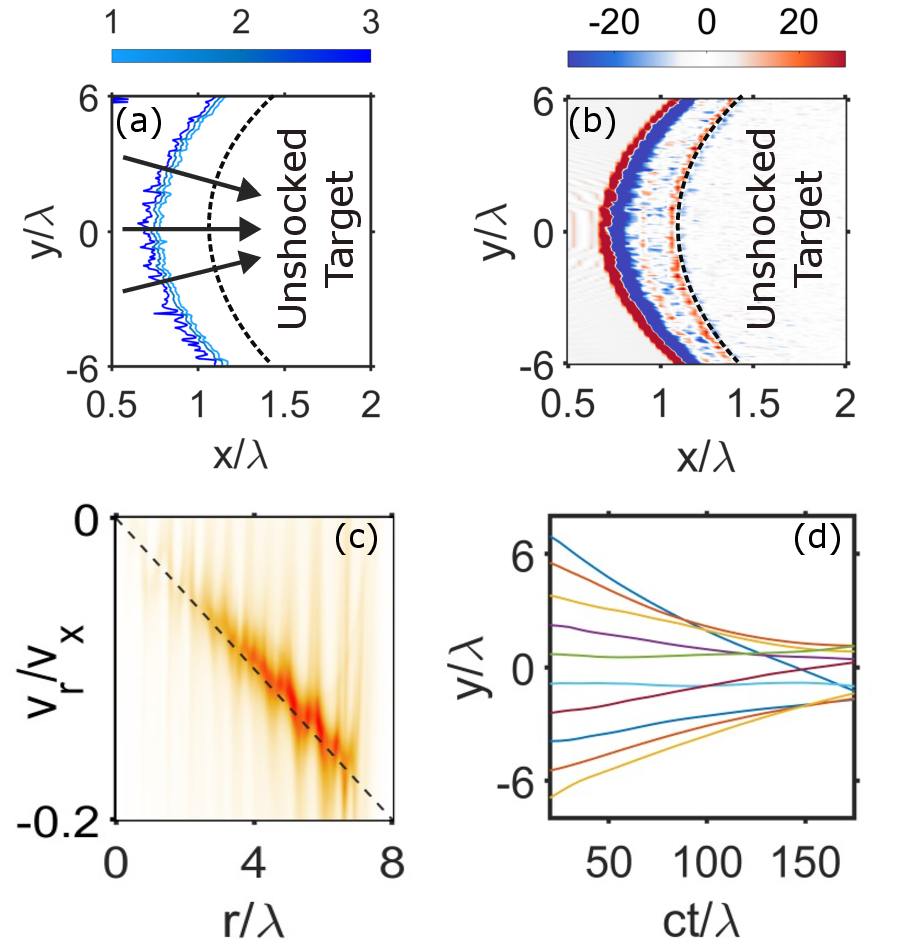}
\caption{Field Structure and ion beam propagation. (a-b) Normalized electric potential $\tilde{\Phi}=-{\Phi_0}
^{-1}\int^x_{\infty} E_x dx=1,2,3$ ,($\Phi_0=E_0\lambda_L=3.2MV$) (a) and net charge density $(n_h-n_e)/n_{\rm crit}$ (b) 16 fs after the laser hits the front surface. Black arrows in (a): direction of ion acceleration, Black dashed lines in (a-b): position of the leading ions that were first accelerated. (c) Ion phase space distirbution at 50fs. Black dashed line: $v_r/v_x=-r/R_c$. (d) Time evolution of transverse ion position versus time for selected ions. Simulation parameters: see table \ref{table:1}.}\label{fig:Field_Trajectory}
\end{figure}

Our simplified model belies several complex effects present in laser-plasma interaction. The surface layer electrons are heated to a finite temperature \cite{Jason_JPP,Jason_PRR,Brunel}, and the surface can undergo transverse instabilities that can perturb the PLI surface \cite{Jason_JPP,Jason_PRR,Vladimir_RTI_2014,Mori_RTI_2015,Mori_RTI_2018,Yan_RTI_2020,Eliasson_RTI_2015,Macchi_RTI_2015} [Fig \ref{fig:3D PIC} (b)]. Furthermore, the quasi-steady propagation of the accelerating shocklike structure is accompanied by small-scale oscillations around the equilibrium velocity of the PLI \cite{Naumova_POP,Robinson_PPCF}. 

To confirm that our model can adequately capture the PLI during the acceleration process, we compare  it with results from Sim I. The electric potential exhibits a parabolic shape 16fs after the laser starts interacting with the target [Fig \ref{fig:Field_Trajectory} (a)]. While there are  perturbations on the electric potential arising from the transverse PLI instability, the potential still accelerates and focuses the ions. The well-localized electric potential coincides with the double layer of electrons moving ahead of ions, with the parabolic shape maintained [Fig \ref{fig:Field_Trajectory} (b)]. The ions accelerated earlier maintain good charge neutrality and the PLI effectively sweeps up the ions and electrons, leaving negligible charge behind. 

As a result of these PLI-generated fields, the accelerated ions gain a transversely focusing velocity dependent on the radial position  [orange, Fig \ref{fig:Field_Trajectory} (c)] at $t_1\approx50\rm fs$ after the end of acceleration process, in good agreement with the prediction given by $v_r/v_x=-r/R_c$  from Eqn \ref{eqn:vion_parax} [black dashed line, Fig \ref{fig:Field_Trajectory} (c)], despite the  perturbations caused by the transverse instability. After propagating through few micron of unshocked material, the ions and the neutralizing hot electrons co-progate during the coasting stage. The hot electron cloud exerts a transverse pressure bending the ion beams away from the axis, evidenced by slight bending of the ion trajectories [Fig \ref{fig:Field_Trajectory}(d)].

\section{Dependence on laser-plasma parameters}\label{sec:expt}

\begin{figure}[t]
\includegraphics[width=0.5 \textwidth]{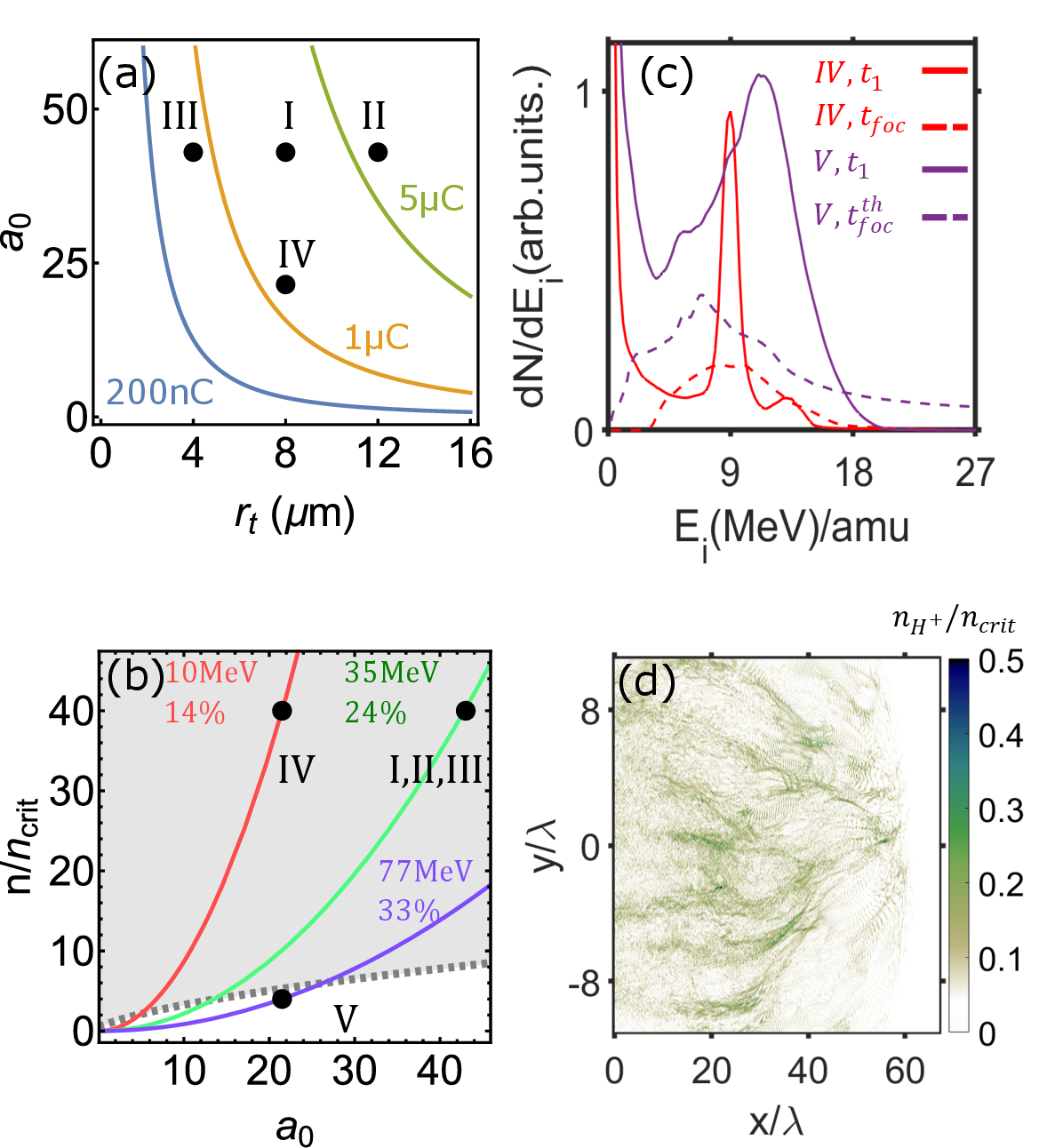}
\caption{Dependence on laser-plasma parameters.  (a) Theoretical estimates for accelerated charge versus target dimension and laser intensity for $t_{\rm laser}=40\mathrm{fs}$ and hydrogen target density $n_i=4.5\times 10^{22}/{\rm cm}^3$. Lines: equi-charge contours from Eqn \ref{eqn:charge}, Black dots: target and laser parameters from Simulations, also see text.  (b) Theoretical ion energy as a function of plasma density $n/n_{\rm crit}$ and laser field strength $a_0$.  Gray dashed line: boundary between classical hole-boring and incomplete hole boring. Colored solid lines: different values of $\Theta=$0.025 (Red), 0.063 (Green), and 0.25 (Purple). Dots: parameters corresponding to simulations .(c) Ion energy spectra for Sim IV and V. Solid Red (Purple) lines: Ion energy spectra after $t_1=50\mathrm{fs}$ for Sim IV (V) . Dashed Red (Purple) lines: Ion energy spectra after $t_{\rm foc}=983\mathrm{fs}$, Sim IV ($t_{\rm foc}^{\rm th}=400\mathrm{fs}$, Sim V). Results for Sim V is multiplied by 10 times for visualization. (d) x-y plane density for Sim V at $t_{\rm foc}^{\rm th}$. See text and table \ref{table:2} in appendix for simulation parameters.
}\label{fig:Theta}
\end{figure}

In our theoretical description, the acceleration stage of CLIA depends only on two independent parameters: focal length $R_c$ and dimensionless parameter $\Theta$; the acceleration dynamics dependent on $\Theta$ such as total charge, energy and conversion efficiency can be estimated using the 1-D HB-RPA model without affecting the the focal length $R_c$. This makes CLIA available to a wide range of laser-plasma parameters without having to compromise $R_c$ or $E_{\rm{k}}^{\rm{ion}}$. We validate these predictions using PIC simulations, and also identify regimes where the model fails to hold.

We first confirm the independence between ion mean energy and focal length $R_c$. We perform two additional simulations with different $R_c=20\mathrm{\mu m}, 60\mathrm{\mu m}$, denoting them Sim II and III, respectively. Here we modify both the target focal length $R_c$ and the target radius $r_t$. Specifically, $r_t=0.2R_c$ such that $r/R_c\ll 1$ holds. The laser intensity, duration, and target density are same as that in Sim I. The mean energy of the ions after the acceleration is close to 35MeV for both simulations, demonstrating that the mean ion energy is independent of focal length. The ions are focused to a much smaller radius than the initial target radius at their respective $t_{\rm foc}$ and $x_{\rm foc}$ for both simulations, showing that indeed the 1-D theory holds independently of $R_c$.

Given the laser peak intensity, the required on-target laser power for planar laser is given by 
\begin{equation}
    P({\mathrm{TW}})=0.087 a_0^2  r_t^2(\mu m)/\lambda_L(\mu m)^2.
    \label{eqn:power}
\end{equation}
 The on-target laser power of Sim II and III are 2.5PW and 40PW, respectively, demonstrating that a wide range of power can be used for CLIA scheme. It is also possible to use lower laser power and intensity, as will be discussed below.

Capitalizing on the validity of 1D-theory for CLIA, we estimate accelerated charge after laser finishes interacting with the PLI. Assuming the laser with duration $t_{\rm L}=40{\rm fs}$ and ion density $n_i=4.5\times 10^{22}/{\rm cm}^3(\rho=0.74{\rm g/cm}^3)$, the accelerated ion charge can be estimated by 
\begin{equation}
    Q\approx |e| n_i \pi r_t^2 d_{\rm accel}\label{eqn:charge}
\end{equation}, where $d_{\rm accel}=c v_{b0}t_L/(c-v_{b0})$ the thickness of the accelerated target and $n_i$ the hydrogen ion density [Fig \ref{fig:Theta} (a)]. The theoretical estimates for the total accelerated charge,  $Q_{\rm th}^{II}=0.7\mathrm{\mu C},Q_{\rm th}^{I}=2.7\mathrm{\mu C},Q_{\rm th}^{III}=6.1\mathrm{\mu C}$ are in qualitative agreement with total ion charge extracted from simulation ($x>5\mu m$ at $t=100{\rm fs}$), $Q_{\rm sim}^{II}=0.5\mathrm{\mu C},Q_{\rm sim}^{I}=2.3\mathrm{\mu C},Q_{\rm sim}^{III}=5.3\mathrm{\mu C}$.

 By changing the laser intensity or plasma density, the mean energy to which the ions are accelerated can be controlled while keeping the focal length same ; since $r_t\ll R_c$ guarantees the applicability of 1D theory, changing the parameter $\Theta$ enables control of mean energy to which the ions are accelerated. As shown in Fig \ref{fig:Theta} (b), the curves characterizing different values of $\Theta$ in the $(n/n_\mathrm{crit}, a_0)$ plane determines the mean energy to which the ions are accelerated. 
 
 To demonstrate this, we perform  Sim IV using the same target as Sim I, but with 1/4th the laser intensity and 2.5PW of laser power incident on the same target. The ion beam is accelerated to $E_i\simeq 9\rm MeV$, close to the theoretically expected value of $10\rm MeV$. The ions propagate with slower velocity, leading to longer $t_{\rm foc}=900\mathrm{fs}$, more than twice that of Sim I. However, the minimal focal spot of the ions is comparable to that of Sim I; lower laser intensity heats the electrons to a lower temperature $T_e \approx 150\rm{keV}$ , almost 1/6th that of  Sim I, and the lower thermal pressure compensates for the longer focal length and focusing time. This leads to comparable focusing as that from Sim I, showing similar widening of energy FWHM. We also note that estimate for accelerated charge still holds in this case [Fig \ref{fig:Theta} (a)], showing qualitative agreement between $Q_{\rm th}^{IV}=1.4\mathrm{\mu C}$ and $Q_{\rm sim
}^{IV}=1.2\mathrm{\mu C}$. 
 
 The scaling in Fig \ref{fig:Theta} (b) may suggest that by keeping the laser intensity same but reducing the mass-density of the target, the ion beam energy can be increased indefinitely. This is true in the ``classical" Hole boring regime where the charge separation can balance the radiation pressure \cite{Naumova_POP}. However, there exists a different acceleration regime for thick targets, where the energy balance, not the momentum balance, governs the Hole-Boring process. This transition to ``incomplete hole-boring" can occur when the electrostatic force cannot balance the radiation pressure. For hydrogen plasmas, this boundary between classical and incomplete hole boring regime is identified by the equation \cite{Incomplete_HB}

\begin{equation}
    n_b\approx 0.618 (1+a_0^2)^{0.314} n_{\rm crit}.
\end{equation}
 For $n_p<n_b$ [below the dashed line in Fig \ref{fig:Theta} (b)], the PLI becomes very unstable unlike in the classical HB-RPA regime, incapable of maintaining a stable shape; the ions are accelerated to higher energy at the expense of beam quality\cite{energy_balance}, and the CLIA scheme becomes inapplicable in this regime.  
 
We demonstrate this with Sim V using a much lower plasma density with $n/n_{\rm crit}=4$ equivalent to a mass density of $7.4\mathrm{mg/cm^3}$ using the same 2.5PW laser. As previously discussed, the radiation pressure cannot be balanced by the electrostatic force from charge separation. The accelerated ion spectra peak is located at a much smaller energy than that predicted by classical theory\cite{Naumova_POP,Naumova_prl,Robinson_PPCF}, with larger energy spread [purple solid line, Fig \ref{fig:Theta} (c)]; the electrons are also heated to a very high temperature, drastically increasing the ion energy spread [purple dashed line, Fig \ref{fig:Theta} (c)]. The formed PLI is unstable and makes the estimate on total charge $Q$ inapplicable; the ions are also not focused, occupying a much larger volume with low density [Fig \ref{fig:Theta} (d)].

Finally, we comment on the power required to demonstrate CLIA. Already from extrapolating the results from Fig \ref{fig:Theta} (b) and power estimate (Eqn. \ref{eqn:power}), on-target intensity of 0.63PW is required for $r_t=4\mathrm{\mu m}$ and $a_0=21.5$, enabling used of a sub-PW lasers for CLIA. We show in appendix that existing PW-class lasers \cite{aleph,BELLA} may be suitable for proof-of-principle demonstration of CLIA (Appendix \ref{sec:ALEPH}).
\section{Discussions}\label{sec:discussion}

Up until now, we limited our discussion to planar lasers (homogenous laser intensity) and hydrogen-only target. We discuss the implications of using lasers with radially dependent transverse profile  (Gaussian, super-Gaussian) and targets with multiple ion species, demonstrating the applicability of CLIA to generate ultra-low emittance multi-ion beams under experimentally practical conditions. 

 \begin{figure*}[t!]
    \includegraphics[width=1\textwidth]{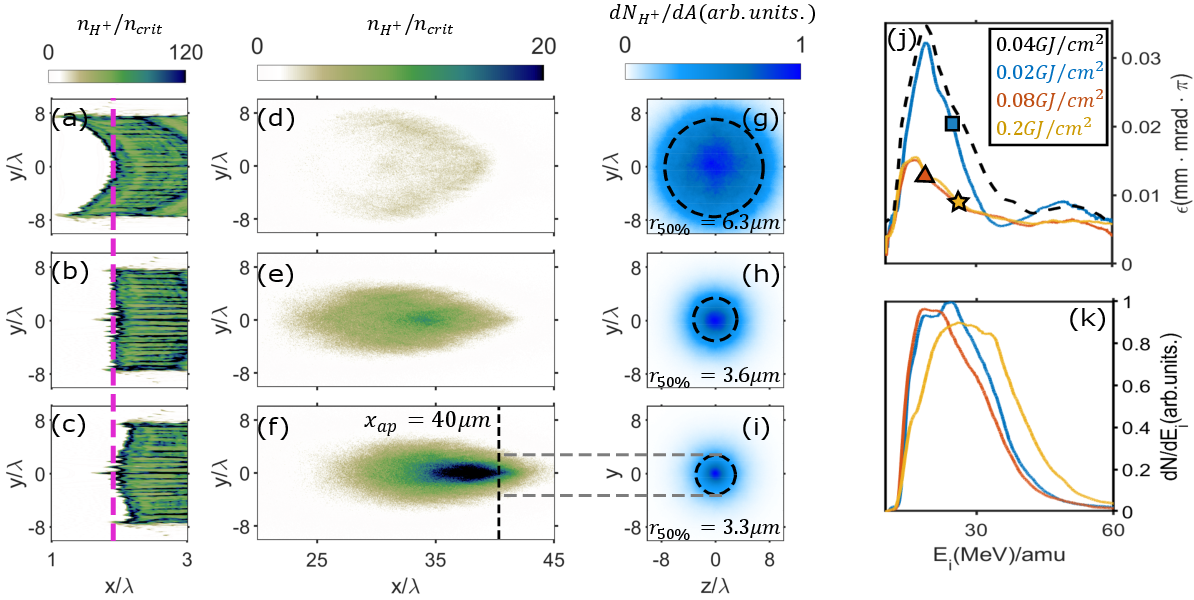}
\caption{Effect of target front shape and  Laser Transverse Profile on CLIA. (a-c): x-y plane density profile for (a) flat target and Gaussian laser (Sim VI), (b) curved target and Gaussian laser (Sim VII), and (c) curved target and super-Gaussian laser (Sim VIII). Pink line: $x_{\rm flip}=1.8\mathrm{\mu m}$.
(d-f): x-y plane density profile at time $t_{\rm foc}=483\mathrm{ fs}$ for Sim VI (d), Sim VII (e), Sim VIII (f) . Black Dashed line: location of virtual screen $x_{\rm ap}$ used in (g-i). Gray Dashed line in (f-i) maps the aperture size in (i) to the aperture location in (f).
(g-i): Number of ions passing the y-z plane at $x_{\rm ap}=40\mathrm{\mu m}$ for Sim VI (g), Sim VII (h), and Sim VIII (i). Dashed lines: contour encompassing 50$\%$ of all ions.
 (j-k): emittance (j) and energy spectra (k) of accelerated ions within $r_\mathrm{50\%}$ for Sim VI (Blue), Sim VII (Red), and Sim VIII (Yellow) at $t_{\rm foc}=483\mathrm{fs}$. Colored shapes in (j): location of peak energy for Sim VI (Square), Sim VII (Triangle), Sim VIII (Star). Dashed line in (j): emittance of accelerated ions for super-Gaussian laser interacting with flat target. Box in (j): peak ion energy flux $J_{peak}=\max{({dE_i/dA})}$ at $t_{\rm foc}$ for respective simulations. Simulation parameters: see table \ref{table:2}.
}\label{fig:superGaussian}
\end{figure*}

\subsection{Effect of Laser and target transverse geometry}\label{sec:Gaussian}

Practically, the available high-power lasers have transversely inhomogeneous profiles ( e.g. Gaussian, super-Gaussian). This effect is especially detrimental in flat targets interacting with Gaussian lasers, increasing the transverse dimension of the accelerated ion beam \cite{Macchi_HB,Zhang_review}. We demonstrate that it is possible to mitigate this effect by using a curved-front target. We also discuss the advantage of using a laser pulse relatively flat intensity profile (super-Gaussian).

We revisit the model in sec \ref{sec:theory} describing the PLI evolution with inhomogeneous laser intensity profiles. The front PLI shape is determined by the  initial target front shape and the laser transverse profile. A Gaussian pulse with amplitude $a(r)=a_0 \exp{(-r^2/\sigma_1^2)} $ has a radially dependent hole-boring velocity. 
 
\begin{equation}
    v^{g}_{b}(r)=\frac{\sqrt{\Theta_0}\exp{(-r^2/\sigma_1^2)}}
    {1+\sqrt{\Theta_0}\exp{(-r^2/\sigma_1^2)}}.
\end{equation}

For an initially flat target with $x(r,0)=0$, the front PLI shape near axis $r\ll \sigma_1$ evolves according to $v^{g}_{b}(r)\approx \sqrt{\Theta_0}(1-r^2/\sigma_1^2)$, and the off-axis elements interacting with lower intensity falls behind, making the PLI shape concave ($dr/dx>0$). Since the ions are always accelerated perpendicular to the PLI, this results in a transversely diverging ion beam.

For a parabolic initial surface, $x(r,0)=r^2/2R_c$, the PLI can be described near axis ($r\ll\sigma_1$) by
\begin{flalign}
    &x(r,t)\approx \frac{r^2}{2 R_c^1(t)}+v_{b0} t,\\  &R_c^1(t)=R_c\left[1-\frac{2R_c v_{b0} t}{\sigma _1^2(1+\sqrt{\Theta})} \right]^{-1}.
\end{flalign}  The quadratic front surface of a target interacting with a Gaussian laser  will initially focus ions to a time-varying focal length $R_c^1(t)$ which increases over time. The PLI become flatter, eventually turning concave at time and distance
\begin{equation}
    t_{\rm flip}=\frac{\sigma_1^2 (1+\sqrt{\Theta})^2}{2c \sqrt{\Theta}R_c}
,
      x_{\rm flip}=\frac{\sigma_1^2 (1+\sqrt{\Theta})}{2R_c}.
\end{equation} Afterwards, the accelerated ions start diverging, limiting the flux of ions that can be focused. 

For a given constant-intensity laser (constant $\Theta$) and fixed laser duration $t_L$, we can also estimate the minimal laser spot-size and the smallest theoretical spot size to which the ions can focus (assuming $T_e\approx 0$). To avoid large focal length $R_c^1(t)$ variation, we can require $t_{\rm push}\equiv\frac{c}{c-v_{b0}} t_{L} \ll t_{\rm flip}$.
This leads to laser spot size requirement for effective focusing:
\begin{equation}
    \sigma_1^2 \gg  2 R_c v_{b0}t_{L}. 
\end{equation} The minimal ion spot size can be also estimated from $R_c^1(t)$, provided $(r\ll \sigma_1)$ holds. For $t_{\rm foc}\gg t_{\rm push}$, the ion focusing can be attributed to variation of PLI curvature $R_c^1(t)$. The ions that were accelerated first and last have a focal length difference, resulting in transverse spread given by
\begin{equation}
    \frac{r_{\rm min}}{r_{t}}= \left(  1-\frac{R_c}{R_c^1} \right) =  \frac{2 R_c v_{b0} t_{L}}{\sigma_1^2(1+\sqrt{\Theta})}. 
\end{equation} As the laser intensity profile becomes flatter ($\sigma_1\rightarrow\infty$), $r_{\rm min}\rightarrow0$, which is why such effects compromising the ion focusing were avoided in our previous simulations using a planar laser.

Alternatively, lasers with super-Gaussian transverse profile can mitigate this effect. For example, a PLI interacting with a 4th order super-Gaussian laser pulse with transverse amplitude written as $a(r)=a_0 \exp{(-r^4/\sigma_2^4)}$. This results in a hole-boring velocity near axis ($z^4\ll\sigma_2^4$),
\begin{equation}
v^{sg}_{b}(r)\approx \frac{\sqrt{\Theta}}{1+\sqrt{\Theta}}+O\left(\frac{r^4}{\sigma_2^4}\right).\label{eqn:sg}
\end{equation}, which is independent of the radial position to a 4th order.  Consequently, the surface does not flip its curvature near axis, enabling acceleration and focusing of a larger flux of ions.

We verify the above predictions using 3D PIC simulations. First, we examine the effect of a Gaussian laser in accelerating a flat target (Sim VI). The on-axis intensity is kept same as that from Sim I: $I=5\times 10^{21} \mathrm{W/cm^2}$, with $\sigma_1=11.3 \mathrm{\mu m}$ corresponding to 10PW of total laser power, with other parameters kept the same. The target is a cylinder with flat front surface at $x=0 \mathrm{\mu m}$ extending to $x=5 \mathrm{\mu m}$ with $r_t=8\mathrm{\mu m}$. After $t_1=50\mathrm{fs}$, the PLI shows a clear concave shape [Fig \ref{fig:superGaussian} (a)] which accelerates diverging ion beams. In comparison, when the same Gaussian laser is incident on a convex front surface as that used in Sim I (Sim VII), the PLI becomes almost flat at $x=1.8\mathrm{\mu m}$ after $t_1=50 \rm{fs}$ [Fig \ref{fig:superGaussian} (b)], in close agreement with $x_{\rm flip}=1.8\mathrm{\mu m}$ and $t_{\rm flip}=45\rm{fs}$. Consequently, all the ions accelerated up to this point will converge, albeit to varying focal lengths, ranging from $R_c=40 \mathrm{\mu m}$ to $R_c=\infty$. Finally, using the same power (10PW) laser with super-Gaussian transverse profile ($\sigma_2=10.1\mathrm{\mu m}$) on the convex target keeps the front surface shape convex [Fig \ref{fig:superGaussian} (c)], and ions accelerated by the PLI are all focused to a  focal length  $R_c\approx 40\mathrm{\mu m}.$

To further quantify the degree of ion focusing, we define a virtual screen located at the theoretical focal point of $x=R_c=40\mathrm{\mu m}$ [Fig \ref{fig:superGaussian} (f)] in the y-z plane. The ion distribution's radial extent in the screen is markedly smaller for the curved targets [Fig \ref{fig:superGaussian} (h-i)].  We define $r_\mathrm{50\%}$ as the radius  that 50$\%$ of the accelerated ions pass through [Fig \ref{fig:superGaussian} (g-i)]. The ``ion count" in the virtual screen shows that using the curved target can focus the ions to almost half the radius for both the Gaussian and super-Gaussian laser [Fig \ref{fig:superGaussian} (h-i)], reducing the $r_{50\%}$ from $6.3\mathrm{\mu m}$ in Sim IV to $3.6\mathrm{\mu m}$ for Sim VII and $3.3\mathrm{\mu m}$ for Sim VIII. Accordingly, the emittance of ions within $r_{\mathrm 50\%}$ are improved for the curved targets. At $t_{\rm foc}=483\mathrm{fs}$, the beam emittances at the peak of energy spectra is largest for  Sim VI with $0.021 \rm{mm} \cdot \rm{mrad} \cdot \pi$, as opposed to Sim VII's $0.012 \rm{mm} \cdot \rm{mrad} \cdot \pi$ and Sim VIII's $0.008 \rm{mm} \cdot \rm{mrad} \cdot \pi$ [Markers, Fig \ref{fig:superGaussian} (g)]. 

While the super-Gaussian laser does improve emittance, and energy flux for curved targets [Fig \ref{fig:superGaussian} (j-k), Fig \ref{fig:appendix2} (c) ], the geometric focusing effect of the curved target plays the main role in reducing the particle emittance. The flat-target results, regardless of the laser profile (Gaussian or super-Gaussian), has a larger minimum focal spot at $x=40\mu m$, with $r_{50\%}=6.3\mu m$ for curved target with Gaussian laser and $r_{50\%}=6.1\mu m$ for the curved target with super-Gaussian laser [Sec. \ref{sec:flatgauss}]. Flat-targets show a consistently larger emittance, as compared to the curved-targets, regardless of the laser shape (Gaussian or super-Gaussian)[Fig \ref{fig:superGaussian}(j)].

\subsection{Multispecies ion target acceleration}\label{sec:multispecies}

\begin{figure}[h]
\includegraphics[width= 0.5\textwidth]{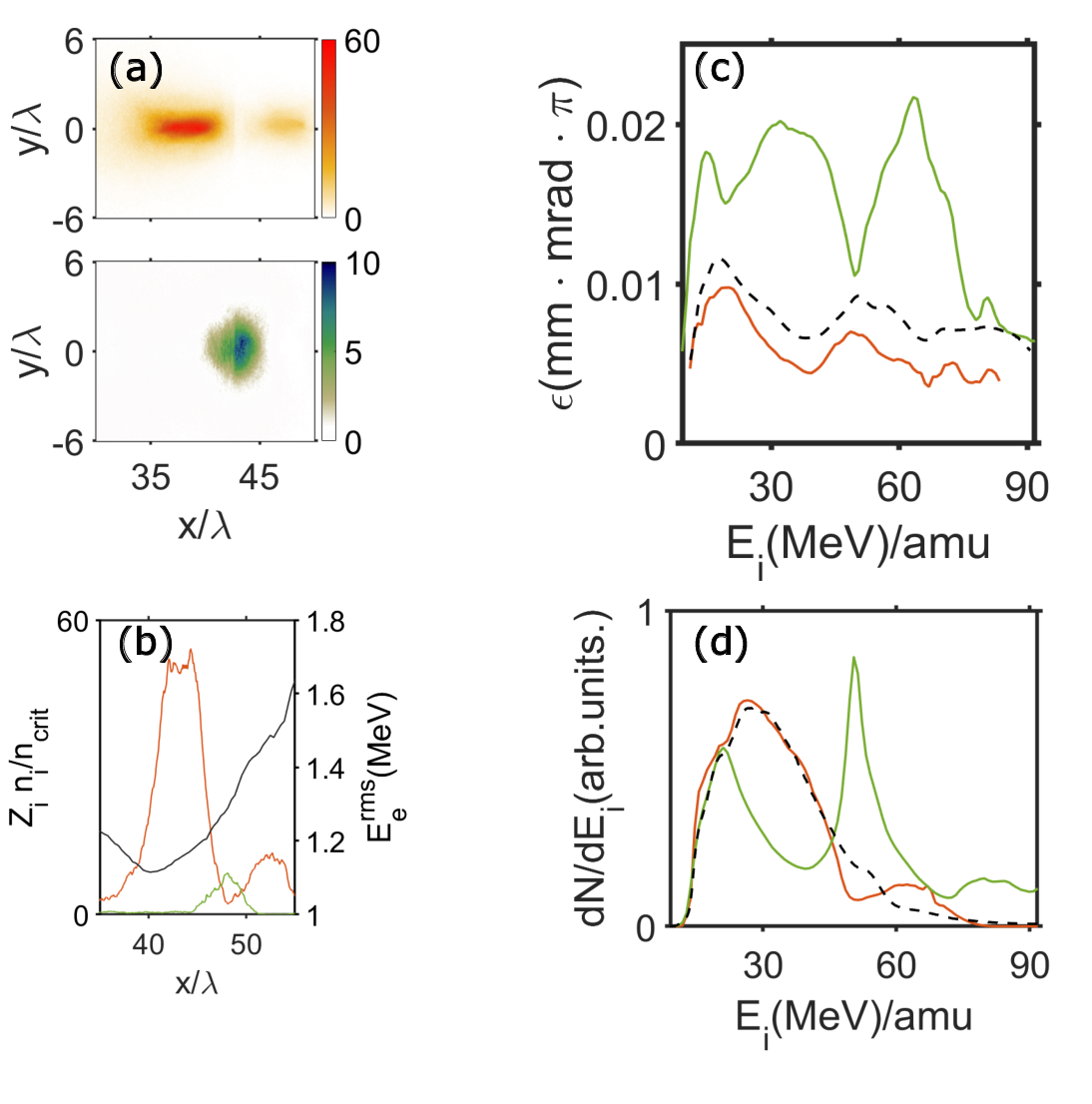}
\caption{Multispecies CLIA simulation results at $t_{\rm foc}=483\mathrm{fs}$ for $\mathrm{CH_{2}}$ target (Sim IX). (a) carbon (Orange) and proton (Green) charge density in x-y plane. (b) On-axis charge density for carbon (Orange) and proton (Green), (left axis), and electron root mean square kinetic energy (right axis) versus longitudinal position. (c-d): Emittance (c) and Energy spectra (d) for ions within through $r_\mathrm{50\%}$ measured at $x_{\rm ap}=40\mathrm{\mu m}$ for carbon (Orange), hydrogen (Green), and hydrogen-only Sim I (Black Dashed). Simulation parameters: On-target power: 10PW, same target shape and mass-density as Sim I, with $\mathrm{C:H=1:2}$, also see table \ref{table:2}.
}\label{fig:multispecies}
\end{figure}

A key feature of HB-RPA scheme is its capability to accelerate multi-species ions to same velocity (or kinetic energy per atomic mass unit) as observed in Ref. \cite{Zhang_review}; shock-like structure propagates through the multispecies plasma, effectively ``snow-plowing" different ions to the same velocity. This makes HB-RPA suitable for applications that require heavier ions \cite{Honrubia_carbon, Honrubia_carbon2,carbon_cancer}. Since the stopping distance for heavier ions have sharper Bragg peak than hydrogen,  these can be a promising alternative for applications such as cancer therapy \cite{carbon_cancer} which can benefit from localized energy deposition. 

On the other hand, the presence of multiple ion species affects the coasting stage of these beams. As shown in single ion species plasma simulations in Sec \ref{section:3D_PIC}, the initially monoenergetic beam energy spectra broadens during the coasting stage due to the thermal pressure of the accompanying electron cloud. In a multi-ion plasma, ions with higher charge-to-mass ratio are more strongly coupled to the hot electron cloud in comparison to ions with lower charge-to-mass ratio  \cite{Kovalev}, consequently gaining/losing energy during the coasting stage.

We verify the above predictions using a target composed with $\mathrm{CH_{2}}$ (Sim IX). By number, the ratio is 1:2 and by mass the ratio is 6:1. The target geometry and density is identical to Sim I, with the charge-density adjusted accordingly, and a planar laser identical to Sim I is used.

Our simulation confirms that both ion species are accelerated to similar  energy per atomic mass units and also gain a focusing transverse velocity distribution similarly to Sim I during the acceleration stage ($t_1=50\mathrm{fs}$), as predicted using the one-dimensional theory. However, the ion species' energy and spatial distribution undergo distinct evolution during the coasting stage; while both species contract transversely, they form distinct distributions longitudinally at $t_{\rm{foc}}$ [ Fig \ref{fig:multispecies} (a), (b)], forming longitudinally separated bunchlets.

 The hot electron cloud neutralizing the ions have a root mean square kinetic energy $E_{e}^{\rm rms}$ dependent on their longitudinal position, with the hotter electrons positioned at the front and rear of the quasi-neutral multispecies ion beam [Fig \ref{fig:multispecies} (b)]; the majority of carbon ions are coupled with the lower kinetic energy electrons at the ``core" region while most of the hydrogen are coupled with more energetic electrons at the periphery.

This position-dependent electron and ion distribution affects both the beam quality and energy; namely, beam quality deterioration of the hydrogen in exchange for additional acceleration/deceleration. To quantify the beam quality, We employ a similar metric as that done in section \ref{sec:Gaussian}, using a virtual screen placed at $x=40\mathrm{\mu m}$. The carbon ions occupy a smaller area on the screen with $r_\mathrm{50\%}^\mathrm{C}\approx 2.4\mathrm{\mu m}$, compared to that of hydrogen $r_\mathrm{50\%}^\mathrm{H}\approx 3.9\mathrm{\mu m}$. 
In line with this varying transverse beam size, the hydrogen emittance in Sim XI is much larger than that from Sim I, while the carbon ion emittance is lower across all energy range [Fig \ref{fig:multispecies} (c)]. The energy spectra of the ion show stark difference, with hydrogen ion energy spectra bifurcating to two  distinct sharp energy peaks at 20MeV and 50MeV  during the coasting stage while the carbon ions energy peak stays at 33MeV/amu, closer to that derived from the $\Theta$ parameter and Sim I. This type of post-acceleration evolution of multispecies plasma ions during the coasting stage may prove beneficial for additional beam property manipulation.

\section{Conclusion and Outlook}

We propose CLIA, a novel ion focusing mechanism in which a Circularly Polarized laser interacting with a thick, opaque target with parabolic front surface simultaneously accelerate and focus a large flux of multi-species ions. The resulting accelerated ion beams' energy and focal length can be tuned independently of each other. The resulting scheme is robust to a wide range of laser-plasma parameters, and is capable of producing ultra-low emittance ion beams.

The produced beams are of broad technological interest. Several of these combined beams may be an energy-efficient way to ignite a dense ICF target via fast ignition \cite{Fernandez}. The ultra-low emittance also makes these beams a promising candidate for cancer therapy \cite{Bulanov_cancer}. The capability of focusing the ion beams in-situ and increasing the ion flux makes these beams promising for many other applications requiring high energy-flux beams focused to a small volume such as generation of warm dense matter \cite{Patel} or neutron production \cite{Pukhov_nat_comm,Roth_neutron}. 

Future work will investigate ways to improve beam quality and ease the laser-plasma restrictions. By reducing the accompanying electron temperature, the beam quality and focusability may be drastically improved \cite{Cooling}. Applicability of elliptically or linearly polarized-laser \cite{Roopendra, ryokovanov} may ease the accesibility of scheme to a wider range of laser facilities. The multispecies ion energy spectra evolution \cite{Kovalev} during the coasting stage provides an additional degree of freedom that may be explored for beam quality manipulation. Finally, the effect of surface curvature on the front-surface trasnverse instability \cite{Mori_RTI_2015,Mori_RTI_2018,Yan_RTI_2020,Vladimir_RTI_2014,Eliasson_RTI_2015} is of both fundamental and practical interest.

 \section*{Acknowledgments}
This work was supported by the National Sicences Foundation under a grant No. PHY-2109087. The authors thank Dr. Vladimir Khudik for helpful feedbacks and the
Texas Advanced Computing Center (TACC) at The University of Texas at Austin for providing HPC resources. All PIC simulations in this paper was performed using the 3D PIC code Smilei\cite{Smilei}. 

\appendix
\section*{Appendices}
\counterwithin{figure}{section}
\counterwithin{table}{section}

\section{Simulation with ALEPH laser parameters}\label{sec:ALEPH}

\begin{figure}[h!]
\includegraphics[width= 0.5\textwidth]{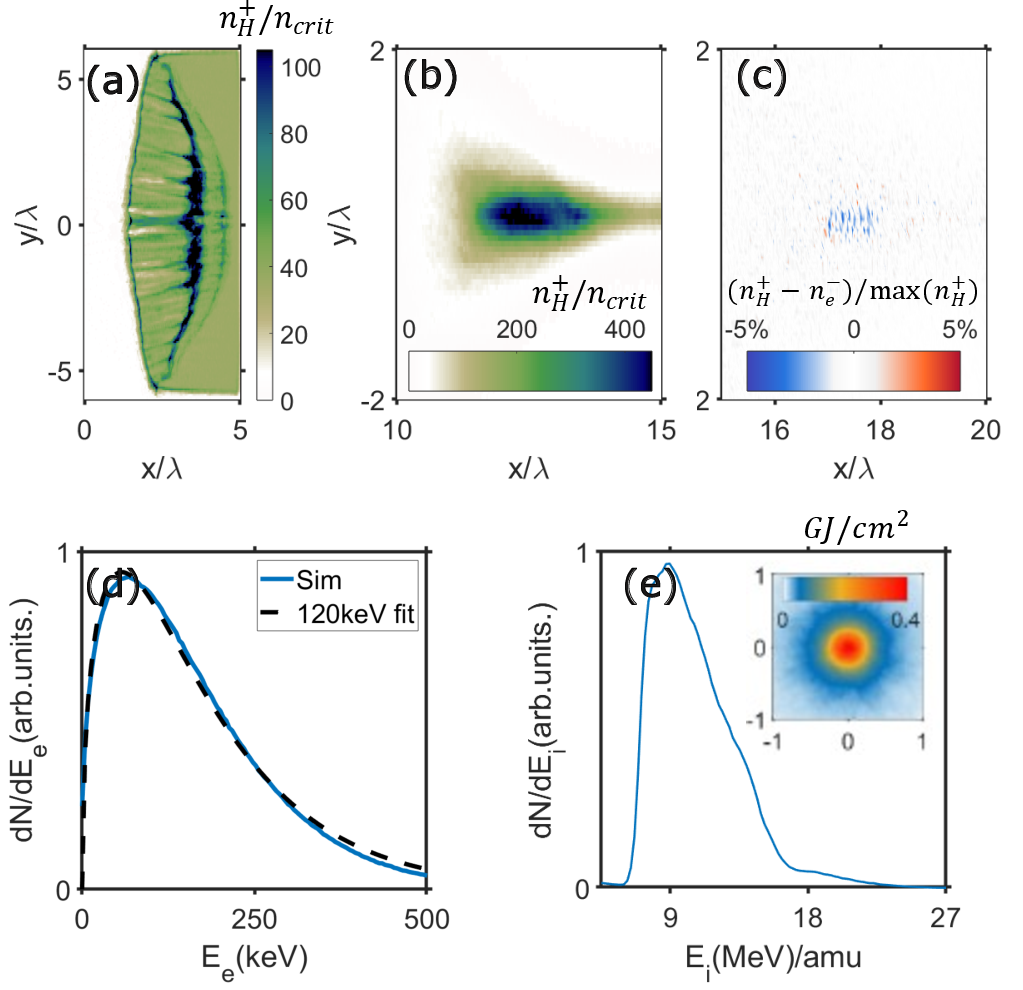}
\caption{ CLIA simulation using ALEPH laser parameters. (a-b) Proton density at $t=66\rm{fs}$ (a) and $t=213\rm{fs}$ (b) in the x-y plane. (c) Charge neutrality $(n_H-n_e)/{n_H^{max}}$ at $t=213$ in the x-y plane. (d) Electron spectra (blue) and Maxwellian fitting (black dashed) at $t=213\rm{fs}$. (e) Accelerated ion spectra and ion energy flux (inset) at $t=213\rm{fs}$
}\label{fig:appendix1}
\end{figure}

We show via PIC simulation that existing PW-class lasers \cite{aleph,BELLA} can be used for proof-of-concept experiments. A Circularly polarized $\lambda_L=0.8\mathrm{\mu m}$, 0.87PW  laser with 30fs flattop temporal profile (corresponding to ALEPH parameter \cite{aleph}) and transversely super-Gaussian profile is incident on a $\rho=0.1g/{\rm cm}^3$ hydrogen target with $R_c=8\mathrm{\mu m}$, $r_t=4.8\mathrm{\mu m}$ with $4\mathrm{\mu m}$ thickness. The laser-plasma parameters are summarized in Table \ref{table:2}. 

After the acceleration has finished, the target front area curvature hasn't flattened, and the converging flow of ions are observed at $x\approx 3\lambda_L$ [Fig \ref{fig:appendix1} (a)]. The ions converge to a radial size of less than a micron at length $x\approx 13\lambda_L=10.4\mathrm{\mu m}$, which is slightly longer than the prescribed focal length of $R_c=8\mathrm{\mu m}$ and increasing in density by almost 13 times, reaching density of $\rho_{max}\approx 1.3g/cm^3$ [Fig \ref{fig:appendix1} (b)] exceeding that of some solids. The plasma is almost completely quasi-neutral with a slight negative charge excess at the peak density region [Fig \ref{fig:appendix1} (c)], in accordance with the quasi-neutral plasma expansion model\cite{Kovalev,Semenov,Mora}. The electron temperature at this point is well-approximated by a Maxwellian fit with $T_e=120\rm{keV}$. The accelerated ion spectra retains a peak at $E_{i}\approx 9\rm{MeV}$, slightly lower than that expected from the 1-D theoretical model $E_i^{th}\approx 13\rm{MeV}$; this is due to the combination of laser transverse profile, target front surface curvature, and coasting stage effect. The ion energy flux at best focus reaches $\mathrm{0.4GJ/cm^2}$
Similar results with higher peak ion energy and flux/density was obtained with simulations using BELLA\cite{BELLA} laser parameters.

\section{Simulation of flat target interacting with a super-Gaussian laser}\label{sec:flatgauss}

We summarize the simulation result of a super-Gaussian pulse interacting with a flat target. The laser parameters are identical to that of Sim VIII, and the target parameters are identical to that of Sim VI. 

\begin{figure}[h!]
\includegraphics[width= 0.5\textwidth]{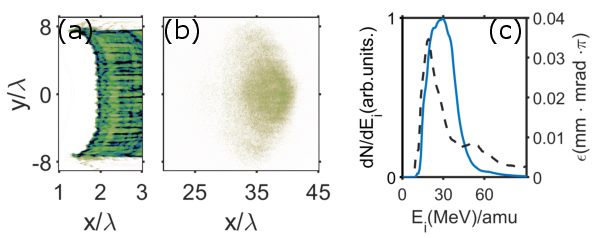}
\caption{ super-Gaussian laser interacting with a flat target. (a-b) Proton density at $t=50\rm{fs}$ (a) and $t=483\rm{fs}$ (b). (c) Emittance (dashed line, right y-axis) and Energy spectra(Blue solid line, left y-axis) versus energy at \rm{483fs} for particles within $r_{50}\%=6.1\rm{\mu m}$ 
}\label{fig:appendix2}
\end{figure}

As predicted from Eqn. \ref{eqn:sg}, the on-axis region of the flat-target stays flat, while the periphery of the target do gain a concave curvature [Fig \ref{fig:appendix2}(a)], leading to a less ``curved" ion beam spatial distribution [Fig \ref{fig:appendix2} (b)] as opposed to SIM VI (flat target, Gaussian laser). This results in slightly lower beam radius passing through $x=40\mu m$: $r_{50\%}=6.1\mu m$ as compared to $6.3\mu m$ of Sim VI. The energy spectra is very similar to that of Sim VIII, but the emittance of the beam is closer to that of Sim VI, showing that the initial target shape has a greater impact on the beam emittance than the laser profile. 

\section{Simulation Details}

\begin{table}[t!]
\caption{\label{table:1} Simulation I details and parameters}
 \begin{ruledtabular}
 \begin{tabular}{l c}
  Variables&  Normalized Values  \\
 \hline
  Plasma density & $n_p/n_{crit}=40$\\
  $a_0=eE_\perp/mc\omega_L$& 43\\
  Longitudinal profile & $\lambda/c$ up/down ramp, $11\lambda/c$ flat\\
  Focal length ($R_c$)& $40\lambda_L$\\
  Target radius ($r_t$)& $8\lambda_L$\\
  Thickness of bulge ($h$)& $0.8\lambda_L$\\
  Thickness at center ($d$) & $5\lambda_L$\\
  Cell size ($\Delta x\times \Delta y \times \Delta z$) & $\frac{\lambda_L}{32} \times\frac{\lambda_L}{16} \times \frac{\lambda_L}{16} $\\
  Particles per cell &  27 \\
  Time step ($\Delta t$) &$\frac{1}{48}\frac{c t}{\lambda_L}$ \\
\end{tabular}
\end{ruledtabular}
\end{table}

We summarize the simulation details for Sim I [Table \ref{table:1}], and outline the relevant parameters for Sims I-IX and the $\Theta$ and $E_i$ from the 1D HB-RPA theory[Table \ref{table:2}]. The required laser powers for demonstrating CLIA can span several orders of magnitude (sub-PW to over 10PW) by using a small radius ($r_t$) target and lower laser intensity (Eqn. \ref{eqn:power}). We  note that the results are applicable to different laser wavelengths; systems such as the ALEPH at 400nm wavelength \cite{aleph} may be relevant for the CLIA scheme, provided that the plasma densities and focal length are scaled accordingly; a carbon-only target with $n/n_{crit}=40$ for such lasers would have mass-density of $0.94g/cm^3$, enabling the use of solid-density targets. 

\begin{table}[h]
\caption{\label{table:2}Summary of simulation parameters. Density $\rho$, Focal length $R_c$, theoretical ion energy, and laser power on target $P_t$ were calculated assuming Laser wavelength $\lambda_L=1\mathrm{\mu m}$ and target transverse radius $r_t=0.2R_c$ except Sim X, where $r_t=0.6 R_c$, $\lambda_L=0.8\mathrm{\mu m}$. Note that for laser intensity Sim I-Sim V assume a flat profile, Sim VI-VII assume a Gaussian profile, and Sim VIII and X assume a transversely super-Gaussian profile. As for the target composition, Sim IX uses a $\rm CH_2$ target. }
 \begin{ruledtabular}
 \begin{tabular}{l c c c c c c}
  Sim&  $R_c(\mu m)$& $a_0$ &$\frac{n}{n_{crit}}$& $\Theta(\frac{E_i}{\rm amu}(MeV))$& $P_t(PW)$ &$\rho(\frac{g}{cm^3})$ \\
 \hline
I& 40& 43& 40& 0.025(35)& 10&0.074\\
II& 60& 43& 40& 0.025(35)& 22.5&0.074\\
III& 20& 43& 40& 0.025(35)& 2.5&0.074\\
IV& 40& 21.5& 40& 0.006(10)& 2.5&0.074\\
V& 40& 21.5& 4& 0.063(77)& 2.5&0.007\\
VI& $\infty$& 43& 40& 0.025(35)& 6.5&0.074\\
VII& 40& 43& 40& 0.025(35)& 6.5&0.074\\
VIII& 40& 43& 40& 0.025(35)& 8.1&0.074\\
IX& 40& 43& 30& 0.025(35)& 10&0.074\\
X& 8& 23.5& 35& 0.09(13.3)& 0.87&0.1\\
\end{tabular}
\end{ruledtabular}
\end{table}

\pagebreak

\pagebreak

\bibliographystyle{unsrt}
\bibliography{ref}%

\end{document}